\documentclass{article}
\usepackage{arxiv}                  
\usepackage[square,numbers]{natbib} 

\usepackage[utf8]{inputenc} 
\usepackage[T1]{fontenc}    
\usepackage{hyperref}       
\usepackage{url}            
\usepackage{amsfonts}       
\usepackage{nicefrac}       
\usepackage{microtype}      
\usepackage{graphicx}
\usepackage{doi}

\usepackage{subcaption}     
\usepackage{enumitem}       

\usepackage{xcolor}
\definecolor{myblue}{HTML}{648fff}
\definecolor{mypurple}{HTML}{785ef0}
\definecolor{myred}{HTML}{dc267f}
\definecolor{myorange}{HTML}{fe6100}
\definecolor{myyellow}{HTML}{ffb000}
\definecolor{lightmyblue}{HTML}{b2cfff}
\definecolor{lightmypurple}{HTML}{c2a9ff}
\definecolor{lightmyred}{HTML}{f08fb0}
\definecolor{lightmyorange}{HTML}{ffb380}
\definecolor{lightmyyellow}{HTML}{ffd480}
\definecolor{lightmygrey}{HTML}{A9A9A9}
\definecolor{mygrey}{HTML}{696969}

\usepackage{tabularray}         
\usepackage{booktabs}           
\usepackage{threeparttable}     
\usepackage{multirow}           
\usepackage{makecell}           

\usepackage{tikz}
\newcommand*\circled[1]{%
  \tikz[baseline=(C.base)]\node[draw,circle,inner sep=0.5pt](C) {#1};\kern-0.015em
}

\usepackage{xspace}     

\newcommand{\gcsnapD}{GCsnap1 Desktop\xspace}
\newcommand{\gcsnapC}{GCsnap2 Cluster\xspace}

\title{Scalable Genomic Context Analysis\\with GCsnap2 on HPC Clusters}
\date{} 
\author{} 

\renewcommand{\shorttitle}{Scalable Genomic Context Analysis with GCsnap2 on HPC Clusters}

\hypersetup{
pdftitle={Scalable Genomic Context Analysis with GCsnap2 on HPC Clusters},
pdfsubject={q-bio.GN, cs.DC},
pdfauthor={Reto Krummenacher,
        Osman Seckin Simsek,     
        Michèle Leemann,
        Leila T.~Alexander,
        Torsten Schwede,
        Florina M.~Ciorba,
        Joana Pereira}
pdfkeywords={Genomic context analysis, Distributed execution, Python, mpi4py, HPC, Performance analysis},
}

\begin{document}
\maketitle

\begin{center}
\vspace{-6em}
    \large{
        Reto Krummenacher\textsuperscript{1},
        Osman Seckin Simsek\textsuperscript{1},        
        Michèle Leemann\textsuperscript{2,3},
        Leila T.~Alexander\textsuperscript{2,3},\\
        Torsten Schwede\textsuperscript{2,3},
        Florina M.~Ciorba\textsuperscript{1}, and
        Joana Pereira\textsuperscript{2,3}  
    } \\
    \vspace{0.5em}
    \normalsize
    \textsuperscript{1}Department of Mathematics and Computer Science, University of Basel, Switzerland\\
    \textsuperscript{2}Biozentrum, University of Basel, Switzerland\\
    \textsuperscript{3}SIB Swiss Institute of Bioinformatics, Basel, Switzerland\\
    \vspace{2.5em}
\end{center}

\begin{abstract}
	\gcsnapC is a scalable, high performance tool for genomic context analysis, developed to overcome the limitations of its predecessor, \gcsnapD. 
    Leveraging distributed computing with mpi4py.futures, GCsnap2 Cluster achieved a 22× improvement in execution time and can now perform genomic context analysis for hundreds of thousands of input sequences in HPC clusters. 
    Its modular architecture enables the creation of task-specific workflows and flexible deployment in various computational environments, making it well suited for bioinformatics studies of large-scale datasets.
    This work highlights the potential for applying similar approaches to solve scalability challenges in other scientific domains that rely on large-scale data analysis pipelines.
\end{abstract}

\keywords{Genomic context analysis \and Distributed execution \and Python \and mpi4py \and HPC \and Performance analysis}

\section{Introduction}
Technological advances, such as high-throughput sequencing and proteomics, have exponentially expanded public biological databases, establishing biology as a data-rich science.
A notable milestone is the sequencing of over 500'000 species, which allowed the life sciences community to identify millions of genes across the tree of life.
Many of these genes encode proteins, the molecular machineries of cells, yet the functions and biological roles of a significant proportion remain unknown, as demonstrated by the Protein Universe Atlas initiative ~\cite{atlas}.
Still, this rapid increase in data poses challenges for large-scale studies~\cite{chaudhari_biological_2024}, as existing methods often struggle to scale efficiently.
One critical challenge lies in genomic context analysis, which studies the genomic neighborhood of a specific protein-coding gene by analyzing which are nearby protein-coding genes that may be associated with the target protein.
When applied across multiple species, comparing such genomic neighborhoods can assist in predicting a protein’s biological function ~\cite{Mavromatis2009} and provide insight into genome structure evolution.
By investigating patterns of gene presence and conservation in these neighborhoods across species, researchers can infer functional associations between proteins, generating hypotheses about their molecular functions and biological role. 

A recently developed tool that supports such studies is GCsnap~\cite{Pereira2021}, here referred to as \gcsnapD.
This open source, freely available tool developed in Python is designed to identify and compare the genomic context of sets of homologous protein-coding genes\footnote{\url{https://github.com/JoanaMPereira/GCsnap}}. 
Starting from a user-provided list of target genes, \gcsnapD retrieves data from multiple public databases, summarizes the collected information, and generates interactive and integrative context visualizations.
While effective for small datasets, \gcsnapD lacks the scalability required to handle more complex workloads. 
Even in a multicore setting, analyzing the context for a single protein-coding gene takes on average 1.66 seconds.
This suboptimal performance is the result of the heavy reliance on online data requests and the inefficient implementation of thread parallelism in \gcsnapD.
One approach to address this limitation is to redesign the application to operate in a distributed high performance computing (HPC) environment.

In this work, we present \emph{\gcsnapC}, a scalable and high performance open-source tool for genomic context analysis on HPC systems, co-designed by computer and life scientists to balance efficiency and usability. 
\gcsnapC offers two key advantages: enabling large-scale genomic context analysis and improving time and energy efficiency compared to executing lengthy scripts on desktop computers. 
By leveraging mpi4py.futures~\cite{rogowski2023}, which showed better suitability over Dask~\cite{Rocklin2015}, 
\gcsnapC scales efficiently from desktops and single nodes to multiple HPC cluster nodes. 
Evaluation results revealed that \gcsnapC delivers robust performance when processing thousands of sequences in a reasonable time. 
In addition, its design highlights the opportunity for bioinformatics tools to become increasingly interoperable and scalable, enabling new discoveries from growing biological datasets.

The remainder of this paper is organized as follows. 
Section~\ref{sec:meth} describes the methodology, including an overview of the GCsnap workflow, followed by an analysis of the execution time of \gcsnapD, the necessary data and the implementation of \gcsnapC, including preliminary tests of the distributed execution frameworks considered.
Section~\ref{sec:results} presents the experimental results of \gcsnapC and includes a detailed discussion of the findings.
Section~\ref{sec:rel_work} reviews existing solutions related to genomic context analysis and studies relying on mpi4py to enable distributed computation.
Section~\ref{sec:conclusion} concludes the paper and provides an outlook for future work.

\section{Methodology}
\label{sec:meth}

This section begins by outlining the general workflow of \gcsnapD followed by the presentation of empirical evidence of its infeasibility for large-scale analysis. 
Then it details \gcsnapC, delineating the required data and providing implementation specifics.
The section ends by describing the frameworks for distributed execution in HPC clusters, which will be used to execute and evaluate \gcsnapC in Section~\ref{sec:results}.

\subsection{\gcsnapD and Its Workflow}

\begin{figure*}[t]
    \centering
    \includegraphics[width=0.9\textwidth, trim={0 6.1cm 1cm 0}]{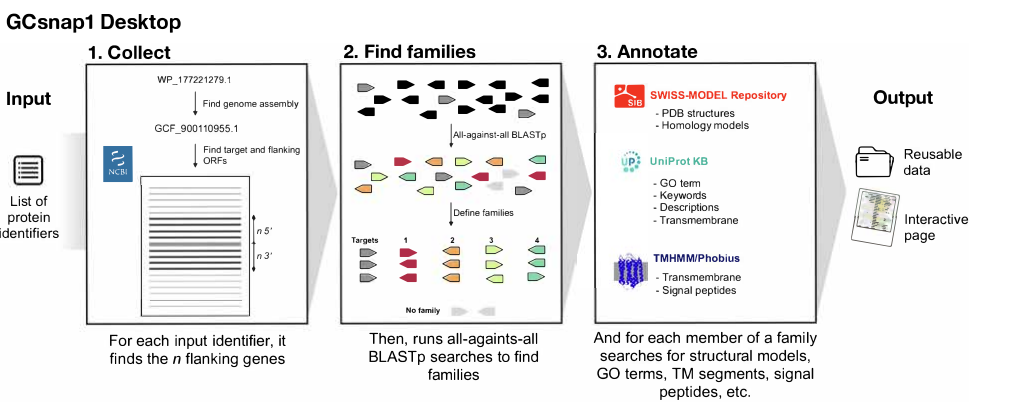}
    \includegraphics[width=0.95\textwidth]{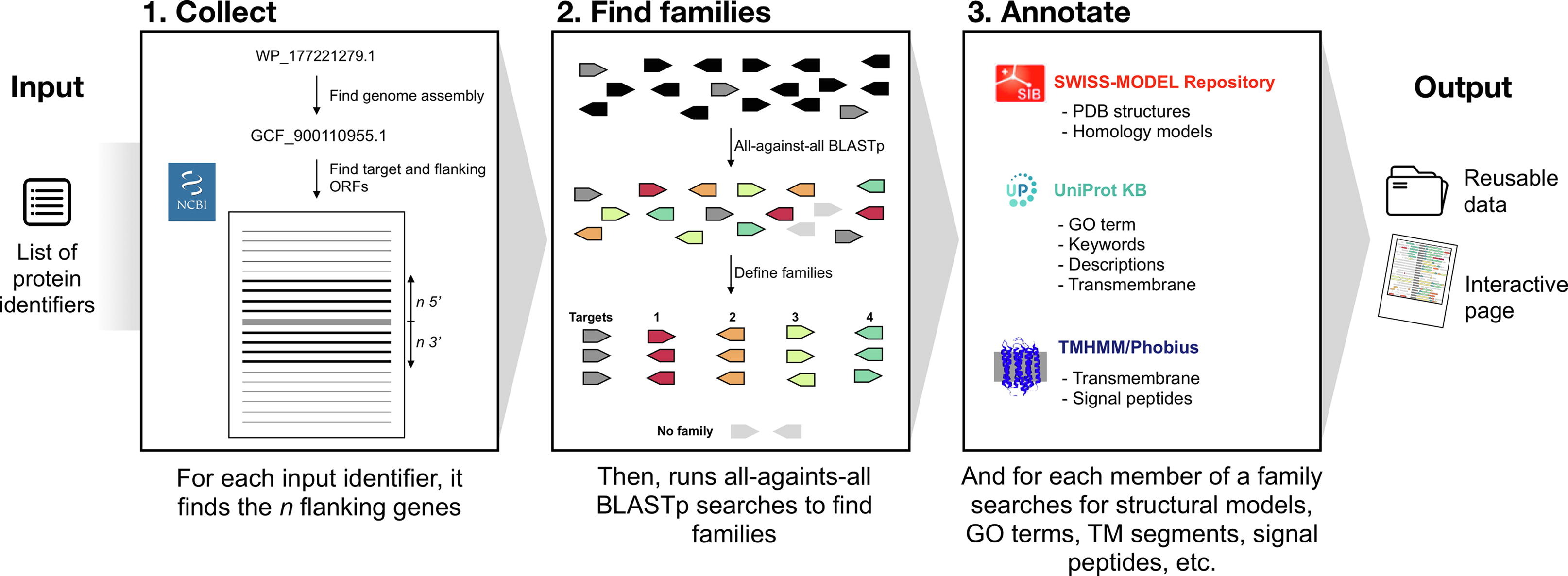} 
    \caption{Schematic workflow of \gcsnapD (from~\cite{Pereira2021}). 
    The input is a list of protein identifiers, referred to as "targets". 
    \gcsnapD comprises 3 steps: 
    (1) Collect: Information is collected from various sources to find the neighboring genes (referred to as flanking genes) of a target. 
    (2) Find families: Gene families within the targets and flanking genes are identified. 
    (3) Annotate: Identified families from the previous step are annotated with functional and structural information. 
    The output consists of annotated gene families and their members stored in text files and presented visually through static and interactive images.}
    \label{fig:gcsnap_workflow}    
\end{figure*}

\gcsnapD is used to facilitate a comparative analysis of the genomic context of genes that encode for evolutionarily related proteins.
It is applicable to both prokaryotic and eukaryotic organisms.
The tool is capable of handling a variety of input formats, collecting data from disparate protein databases, searching for homologous proteins, identifying clusters, and synthesizing this information into static and interactive plots.
The organizations responsible for providing and managing the vast amount of data include the National Center for Biotechnology Information (NCBI) providing the Genetic Sequence Database (GenBank GCA) and the Reference Sequence Database assembly accesses (RefSeq GCF), the Universal Protein Resource Database (UniProt), the European Molecular Biology Laboratory (EMBL), and the Swiss Model Repository (SMR), from which protein structure information is retrieved.
The various tasks performed by GCsnap form a workflow, illustrated in Figure~\ref{fig:gcsnap_workflow}, and can be divided into three workflow steps: collect, find families, and annotate. 

\gcsnapD~begins by processing either a list of sequence identifiers or text files containing sequence identifiers from disparate protein ID standards.
It then identifies the genome assemblies where the target proteins are annotated and extracts neighboring genes situated in proximity to their encoding gene, which are called flanking genes.
Step 2 is responsible for finding the protein families.
The similarities of protein sequences encoded by all collected genes are computed based on an all-against-all sequence similarity search using BLASTp~\cite{Blastp} or MMseqs2~\cite{MMseqs_paper}, followed by clustering them into families.
In Step 3, GCsnap annotates families with structures and function-related terms, as well as family members with sequence features, such as signal peptides or transmembrane segments~\cite{phobius,tmhmm} and taxonomy information from various repositories and online tools. 
The information collected is stored in text files and presented visually through static and interactive images.
More details can be found in Pereira~\cite{Pereira2021}.

\begin{figure*}[t]
    \centering   
    \includegraphics[width=1\textwidth,trim={0.7cm 0 0.6cm 0},clip]{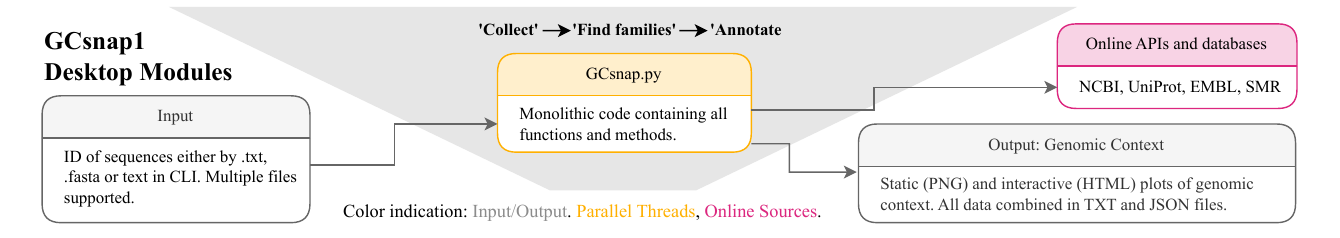}
    \caption{Software modules of \gcsnapD workflow (depicted in Figure~\ref{fig:gcsnap_workflow}), showcasing the three workflow steps consolidated within the monolithic GCsnap.py file. 
    }
    \label{fig:modularization_gcsnap1}
\end{figure*}

\gcsnapD is designed and implemented in a straightforward manner, in a single Python module as illustrated in Figure~\ref{fig:modularization_gcsnap1}.
\texttt{GCsnap.py} is a comprehensive Python script comprising 3'427 lines of code (LOC), targeted for desktop/single-node computing. 
Runtime arguments are specified via the command line. 
The code employs Python's multiprocessing module with threads to facilitate node-level parallelism. 

\begin{figure*}[!ht]
    \centering
    \includegraphics[width=0.6\linewidth]{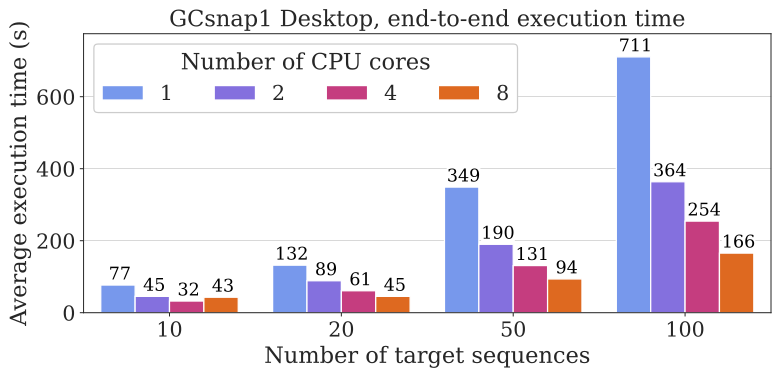}  
    \caption{Average end-to-end execution time of \gcsnapD over 5 repetitions with different numbers of CPU cores and input targets. 
    The number of multiple threads within the main Python process is equal to the number of CPU cores.
    Tests conducted on an AMD EPYC 7742 CPU with 64 CPU cores. 
    }
    \label{fig:gcsnap1_overview}
\end{figure*}

\subsection{\gcsnapD Assessment} \label{sec:assessment}
The current implementation of \gcsnapD suffers from a significant performance limitation. 
Figure~\ref{fig:gcsnap1_overview} illustrates the average execution time for varying configurations. 
The runtime arguments were set to generate results comparable to \gcsnapC (described later in Section~\ref{sec:gcsnapC-implementation}). 
The most salient observation is the considerable end-to-end execution time. 
Even though \gcsnapD uses thread parallelism, the average time required to analyze the genomic context of 100 targets is 166 seconds when running on 8 CPU cores which is sub-par in terms of both productivity and performance.

This poor performance can be attributed to the Collect workflow step, where a large number of queries is made to public databases and on-line servers. 
The speed of the internet connection and the efficiency of the online application programming interfaces (API) in handling queries are the limiting factors in such operations. 
Furthermore, \gcsnapD's thread parallelism offers limited performance due to threads in Python being bound by the Global Interpreter Lock (GIL). 
The GIL mechanism of the CPython interpreter ensures that only one Python thread executes the bytecode at a time, even on multi-core machines. 
While this offers protection against concurrent accesses, it limits the achievable performance gain, as Python threads do not run in parallel. 
Moreover, there is a general problem when issuing API requests from multiple threads. 
As more and more threads send requests via the APIs, the API limits are reached sooner, causing more requests to block and increasing the number of retry attempts.
The assessment supports the claim that \gcsnapD cannot manage substantial workloads consisting of thousands of genome sequences. 

\subsection{\gcsnapC Parallelism and Data}\label{sec:data}
To enable distributed execution of \gcsnapC, we need to leverage data parallelism in workflow Steps 1 (Collect) and 3 (Annotate). 
Considering Step 2 (Find families), the only parallelizable part is the assignment of identified families to flanking genes.
The other tasks involved in this step require different approaches.
The process of determining families involves the computation of all-against-all sequence similarities, followed by clustering of the resulting distance matrix, which contains all sequences of interest.
The former can be efficiently executed with MMseqs2~\cite{MMseqs_paper}, while the latter relies on SciPy's implementation of hierarchical clustering~\cite{scipy}, which is based on the single linkage clustering algorithm~\cite{muellner}.

\begin{table*}[!b]
    \small 
    \centering
    \caption{Data necessary to execute \gcsnapC (as retrieved in March 2024).}
    \label{tab:data}
    \begin{tblr}{width=\textwidth, colspec={Q[l,m,wd=50mm]|Q[r,m,wd=14mm]|Q[r,m,wd=18mm]|Q[l,m,wd=21mm]|Q[l,m,wd=40mm]},stretch=1.2} 
        \textbf{Data} & \raggedright \textbf{File count} & \textbf{Total Size}  & \textbf{Stored as} & \textbf{Source} \\ 
        \hline
        UniProt mapping between ID standards & 1 & 44 GB & mappings.db & \href{https://ftp.uniprot.org/pub/databases/uniprot/current_release/knowledgebase/idmapping/idmapping_selected.tab.gz}{idmapping\_selected.tab.gz} \\ 
        \hline
        NCBI summary table GenBank assemblies\textsuperscript{a} & 1 & 991 MB & TXT & \href{https://ftp.ncbi.nlm.nih.gov/genomes/genbank}{assembly\_summary\_genbank.txt} \\  
        \hline   
        NCBI summary table RefSeq assemblies\textsuperscript{a} & 1 & 160 MB & TXT & \href{https://ftp.ncbi.nlm.nih.gov/genomes/refseq}{assembly\_summary\_refseq.txt } \\  
        \hline    
        \makecell[l]{GenBank assemblies\textsuperscript{a}\\ (\_genomic.gff.gz)} & 1'678'176  & \makecell[r]{461 GB \\ min: $<$1 KB \\ avg: 274 KB \\max: 188 MB} & GFF.gz & \href{https://ftp.ncbi.nlm.nih.gov/genomes/all/GCA}{Index of /genomes/all/GCA} \\  
        \hline   
        \makecell[l]{RefSeq assemblies\textsuperscript{a}\\(\_genomic.gff.gz)} & 362'140 & \makecell[r]{138 GB \\ min: $<$1 KB\\ avg: 380 KB \\ max: 79 MB} & GFF.gz & \href{https://ftp.ncbi.nlm.nih.gov/genomes/all/GCF}{Index of /genomes/all/GCF} \\  
        \hline   
        \makecell[l]{GenBank sequence files\\ (\_protein.faa.gz)}& 1'676'631 & \makecell[r]{1.29 TB \\ min: $<$1 KB\\ avg: 770 KB\\ max: 304 MB)} & sequences.db & \href{https://ftp.ncbi.nlm.nih.gov/genomes/all/GCA}{Index of /genomes/all/GCA} \\  
        \hline   
        \makecell[l]{RefSeq sequence files\\ (\_protein.faa.gz)}& 1'676'631 & \makecell[r]{289 GB \\ min: $<$1 KB\\ avg: 780 KB\\ max: 29 MB)} & sequences.db & \href{https://ftp.ncbi.nlm.nih.gov/genomes/all/GCF}{Index of /genomes/all/GCF} \\  
        \hline   
        NCBI taxonomy (new\_taxdump.tar.gz)& 1 & 324 MB & rankedlineage.dmp & \href{https://ftp.ncbi.nih.gov/pub/taxonomy/new_taxdump}{new\_taxdump.zip} \\
        \hline
    \end{tblr}
    \begin{tablenotes}
        \item \hspace{-12pt} \textsuperscript{a} Used to create mapping between GenBank/RefSeq IDs, assembly accessions, assembly files, and taxonomy ID stored in assemblies.db.
    \end{tablenotes}
\end{table*}  

Since compute nodes in HPC clusters are typically not connected to the Internet, they cannot query online APIs. 
Therefore, it is imperative that all data is available locally to \gcsnapC.
The necessary data is described in Table~\ref{tab:data}, along with the provenance and storage methods.
\gcsnapC processes two different data formats: 
(1) \textit{raw files} are parsed to extract relevant information, and 
(2) \textit{databases} are used to emulate APIs for quick data retrieval during execution.
Raw files can be downloaded directly, e.g., the assembly files from which the flanking genes are extracted, whereas databases must first be created. 
In some cases, there is a single source for both data formats. 
For example, UniProt provides a single file containing all mappings between ID standards; NCBI does not provide such a combined summary for the mapping from GenBank/RefSeq ID to assembly accession.
We generate this mapping by extracting the information from the raw files and store the mapping in a database. 
Similarly, coding sequences are parsed from available protein sequence files. 
We use the SQLite database with its Python implementation because it is a lightweight solution that does not require running a separate server~\cite{sqlite}, thus minimizing software dependencies.

The above approach works when the data underlying the APIs is freely available for download. 
This is true for most of the sources used by \gcsnapC, with certain exceptions.
For example, information from EMBL can only be retrieved via their API, making pre-downloading impossible.
A similar limitation exists for transmembrane segment and signal peptide prediction tools.
To overcome this situation, \gcsnapC allows users to provide such files to be used for annotation in Step 3.

\subsection{\gcsnapC Implementation} \label{sec:gcsnapC-implementation}
\begin{figure*}[!htb]
    \centering 
    \includegraphics[width=0.91\textwidth,trim={0.6cm 0.9cm 0.5cm 1cm},clip]{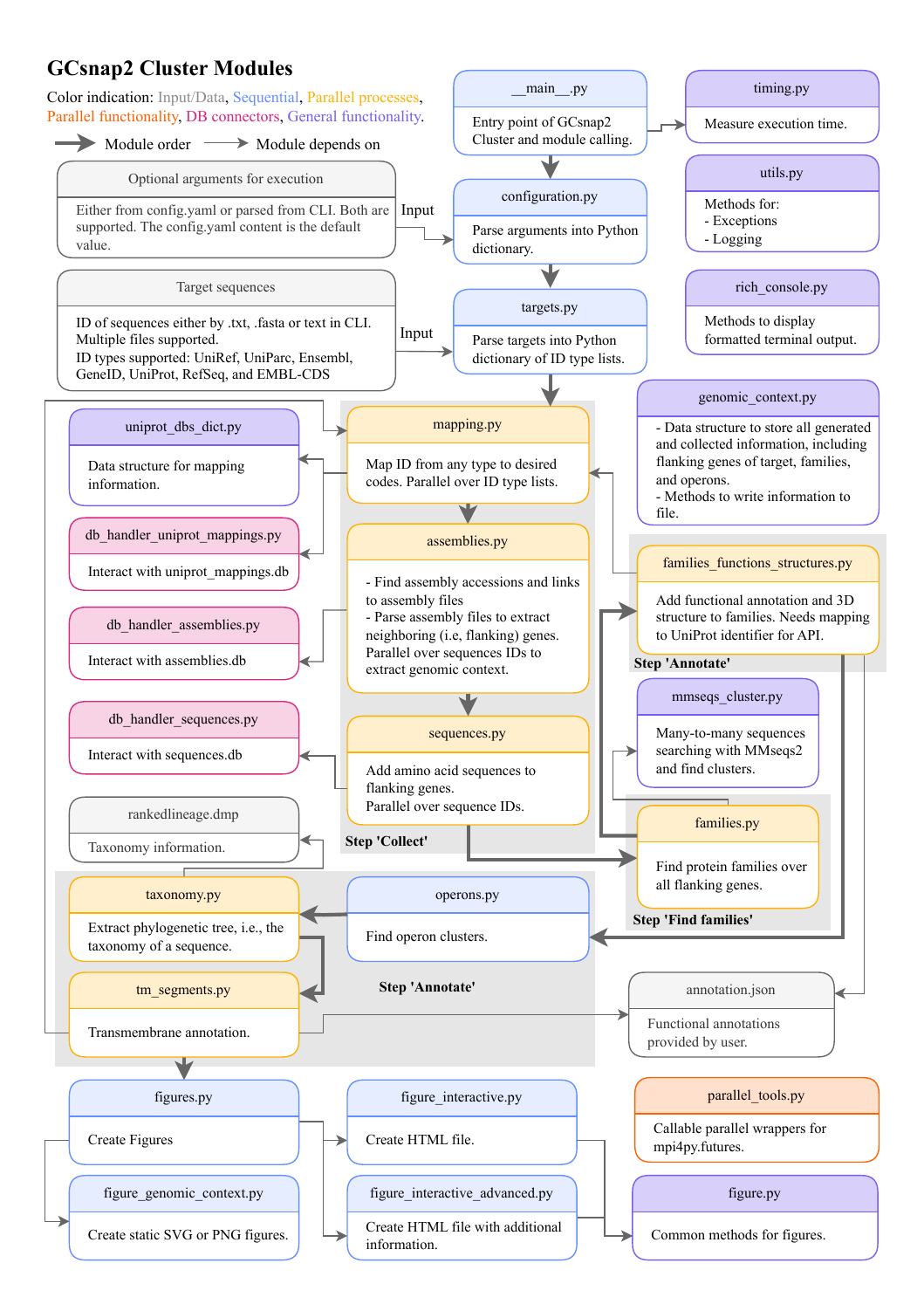}
    \caption{Software modules and dependencies of \gcsnapC. Modules without arrows are used in most other modules and dependencies are not shown for simplicity.}
    \label{fig:modularization_cluster}
\end{figure*}

\gcsnapC is written in Python and is freely available open-source\footnote{\url{https://github.com/GCsnap/gcsnap2cluster}, v1.0.0}.
The tool comprises 26 software modules with 6'794 LOC.
Figure~\ref{fig:modularization_cluster} shows these modules, their dependencies and the program flow. 
The modules corresponding to the three workflow steps (of Figure~\ref{fig:gcsnap_workflow}) are highlighted with gray-shaded boxes.
This modular software design allows for module-level optimization without disrupting the tool's use in production and for future reuse of components in other projects. 
Next, we provide a brief overview of the modules.

The \textbf{\texttt{\_\_main\_\_.py}} module controls the execution by calling other modules and recording the execution time using the functionality provided by \texttt{timing.py}. 
First, the run-time arguments passed by the user either via the command line or via a configuration file are parsed by \texttt{configuration.py} before \texttt{targets.py} extracts the provided targets and determines which ID standard each target belongs to.
The modules "extensive logging" (\texttt{utils.py}) and "formatted terminal output" (\texttt{rich\_console.py}) provide the user with information about the execution state of \gcsnapC.
All information collected in the following steps is stored in an instance of \texttt{genomic\_context.py}.

The \textbf{collect} step relies on the downloaded data and the previously created databases (Section~\ref{sec:data}).
The retrieval of information is enabled by three database handlers (\texttt{db\_handler\_*.py}), shown in pink. 
Collect consists of three substeps:
(1) the module \texttt{mapping.py} which maps all target sequence IDs to their corresponding NCBI IDs, which is used as the unique identifier in assemblies.db and sequence.db. 
(2) \texttt{assemblies.py} fetches all assembly accessions and file names that contain the assembly information followed by the extraction of the flanking genes of the targets.
(3) the module \texttt{sequences.py} queries the amino acid sequence of all targets and flanking genes.

In the \textbf{find families} step, based on previously retrieved amino acid sequences,  \texttt{families.py} uses MMseqs2 and SciPy's hierarchical clustering (\texttt{mmseqs\_cluster.py}) to identify families within the targets and flanking genes. 

In the \textbf{annotate} step, families and their members are annotated, in substeps, with: (1) functional and structural information \texttt{families\_functions\_structures.py}, (2) sequence features such as signal peptides and transmembrane segments (\texttt{tm\_segments.py}), (3) operons that are identified via clustering (\texttt{operons.py}), (4) the phylogenetic tree (\texttt{taxonomy.py}).
Substep (4) relies on the previously downloaded data, while substeps (1) and (2) use user-provided input.

Of particular significance is the provision of parallelization of the modules, shown in yellow, through a callable wrapper function class (\texttt{parallel\_tools.py}), indicated in orange. 
In essence, this wrapper accepts a list of elements and a function as input, subsequently distributing the function across the desired computing resources on an HPC computing cluster.
We considered Dask and mpi4py as a foundation for the wrapper function, and we describe them in the next section.

\subsection{Distributed Execution Frameworks} \label{sec:distributed_execution}
Dask is a flexible task-based parallel computing library introduced by Rocklin~\cite{Rocklin2015}.
It encodes task graphs by using Python data structures, specifically dictionaries, tuples, and callables.
In these task graphs, the nodes represent individual tasks and the edges represent the dependencies of the task.
The \emph{Dask.jobqueue} module~\cite{dask} provides an API for interacting with HPC cluster resource and job management systems (RJMS), such as SLURM.
The module enables the direct request of resources on an HPC system from within the Python programming language.
At the system level, this process begins with the submission of Python jobs to the HPC cluster. 
Once resources are allocated by the RJMS to the Python job, the main Python thread executes the Dask.jobqueue scheduler, which in turn requests additional compute nodes to execute the Python task graph.
Essentially, this workflow involves two resource requests and allocations: the first occurs when resources are allocated through the job submission script, and the second is initiated by the Dask.jobqueue scheduler to manage the task graph’s execution across the compute nodes. 

The Message Passing Interface (MPI)~\cite{mpi} is a standard API that enables fast information exchange between processes in distributed systems.
In MPI, the term "world communicator" refers to the collection of processes that constitute a single process group for synchronization and data exchange.
All MPI processes (also called ranks) are by default members of the global world communicator.
The \emph{mpi4py} Python library~\cite{Dalcin2021} provides bindings that enable direct access to MPI API calls from a Python script.
The development of mpi4py started in 2004 and follows the evolution of the MPI standard~\cite{mpi}.
Mpi4py enables the transfer of Python objects, such as NumPy arrays, through a variety of methods, including nonblocking MPI communication primitives.
The user-friendly interface of mpi4py provides significant ease of use. 
However, users must still adhere to the coordinator-worker pattern employed in distributed memory parallel programming. 
In this pattern, rank 0 typically acts as the coordinator, while all ranks act as workers.
In response to the need of the Python user community for a more straightforward approach to deploying computations on HPC clusters, mpi4py introduced the \emph{mpi4py.futures} module~\cite{rogowski2023}, which can be used through the \emph{MPIPoolExecutor}~\cite{mpi4pyfutures}. 

\begin{figure}[t]
    \centering
    \includegraphics[width=0.6\linewidth, trim={2cm 0cm 0cm 0cm}, clip]{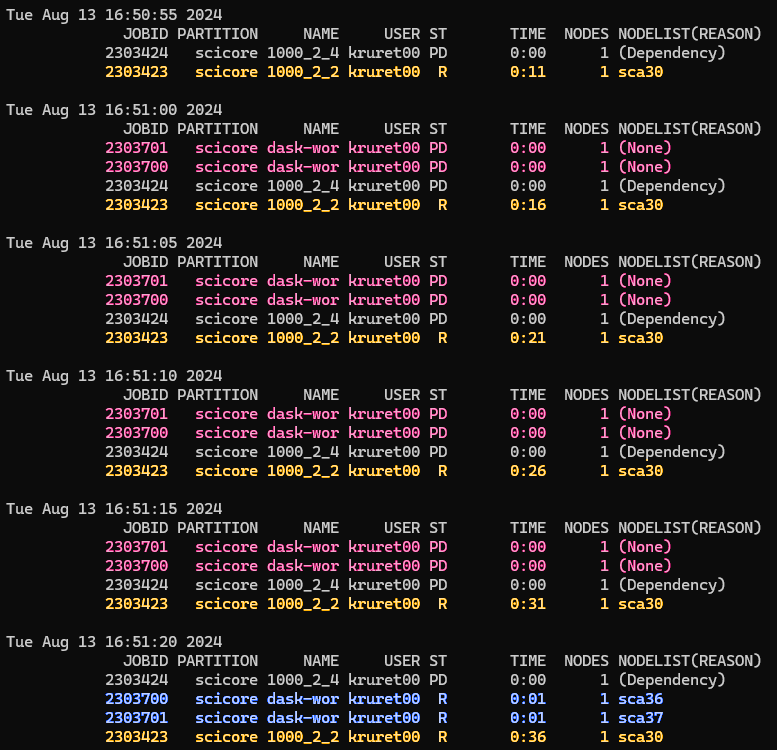}
    \caption{Dask.jobqueue execution on an HPC cluster. 
    The main Python thread \textcolor{myyellow}{\texttt{1000\_2\_2}} executing the Dask.jobqueue scheduler runs (\textcolor{myyellow}{\texttt{R}}) on its assigned computing node \textcolor{myyellow}{\texttt{sca30}} and requests allocation of additional nodes for two Dask-workers (\textcolor{myred}{\texttt{dask-wor}}). 
    Here, the main thread \textcolor{myyellow}{\texttt{1000\_2\_2}} runs without progress for 19 seconds (0:16-0:35) until the two pending (\textcolor{myred}{\texttt{PD}}) \textcolor{myred}{\texttt{dask-wor}} threads are allocated computing nodes (\textcolor{myblue}{\texttt{sca36}}, \textcolor{myblue}{\texttt{sca37}}) and start execution (\textcolor{myblue}{\texttt{R}}) as orchestrated by the Dask.jobqueue scheduler.}
    \label{fig:sc_nodes_running}
\end{figure}

\begin{figure}[t]
    \centering 
    \includegraphics[width=0.6\linewidth]{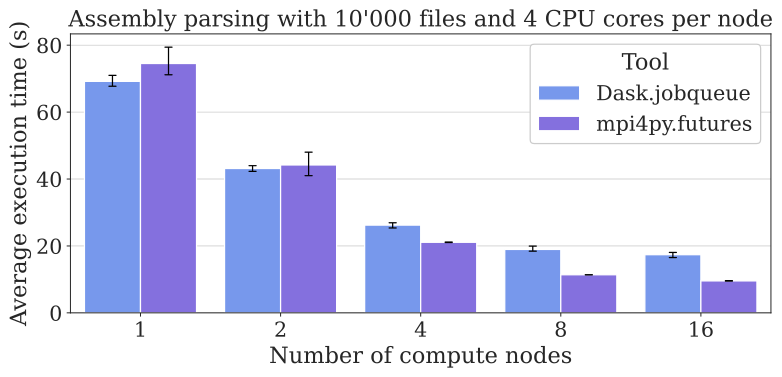}
    \caption{Average execution time of the assembly file parsing test with Dask.jobqueue and mpi4py.futures over 5 repetitions with different numbers of nodes, 4 CPU cores per node, and 10'000 files. 
    The number of total processes with Dask.jobqueue and mpi4py.futures equals the number of compute nodes $\times$ the number of CPU cores per node.
    Error bars represent the minimum and maximum of 5 repetitions. 
    Experiments conducted on various nodes with 2 Xeon E5-2640v4 CPUs and 20 CPU cores.}
    \label{fig:experiments_dask_mpi}
\end{figure}

Before examining the experimental results in Section~\ref{sec:results}, it is important to outline here key observations regarding the implementation of Dask and mpi4py.futures on an HPC system.
 Python scripts are initiated by submitting a job request that specifies the cluster resources required for execution.
The distinction between the aforementioned tools lies in the sequence of resource requests.
As previously stated, Dask.jobqueue issues requests for additional worker nodes as required by the Python script.
Consequently, the primary Python thread, which executes the Dask.jobqueue scheduler, is initiated with a job script that requests one node to the RJMS, and Dask handles the subsequent steps.
This differs from mpi4py.futures, where all necessary resources are requested to the RJMS only via the job script. 
Moreover, Dask requires an additional node to facilitate the operation of the main thread and of the Dask scheduler.

The operational scenario described above is illustrated in Figure~\ref{fig:sc_nodes_running}, where a Python thread that executes the Dask.jobqueue scheduler initially runs on one node and, after a delay, additional worker nodes are allocated to execute tasks, without a priori visibility to the RJMS of the total number of nodes required by a Python job using Dask. 
This approach appears suboptimal, especially in the context of HPC systems commonly used by life scientists, where resources are in high demand, and job queues often involve significant waiting times.
A user must wait twice to obtain the required resources: once in the system job queue and once more within the Python+Dask job.
During the second waiting phase, the resources of the node on which the main Python thread runs without progress are also blocked.
One potential solution to this issue is to manually initiate the Dask.jobqueue scheduler and Dask workers from the same job script submitted to the RJMS, thereby circumventing the need for the Dask.jobqueue scheduler to be started from within the Python script.
Nevertheless, this approach would not solve the issue of not fully using the node executing the main Python thread with the Dask scheduler.

mpi4py.futures has recently been shown to consistently outperform Dask.jobqueue in most scenarios~\cite{rogowski2023}.
Therefore, we tested these two approaches with \gcsnapC for opening and parsing a variable number of assembly files, selected from a subset of the total downloaded files.
The primary findings are illustrated in Figure~\ref{fig:experiments_dask_mpi}.
In the single-node test, Dask.jobqueue shows superior performance, whereas mpi4py.futures exhibits a lower average execution time with more than two computing nodes.
No further improvement was observed beyond eight computing nodes.
This is due to the relatively limited workload size, which consists of 10'000 files to be opened and parsed.
Both Dask.jobqueue and mpi4py.futures exhibit significant coordination and communication overhead, which dominates the computation of such a small workload.
Given that the design goal for \gcsnapC is to perform well across many computing nodes, we use mpi4py.futures, the superior choice compared to Dask.jobqueue, in \gcsnapC.
We evaluate its performance and describe the results in Section~\ref{sec:results}.

\section{Results and Discussion} \label{sec:results}
This section presents the experiments conducted to evaluate \gcsnapC with mpi4py.futures with sequences from dark protein clusters, classified in the Universal Protein Atlas~\cite{atlas} and discusses the results.
To assess the impact of several parameters, we conducted three sets of factorial experiments with various ranges and combinations of parameters, comprehensively summarized in Table~\ref{tab:experiments}.
We measure performance with the average parallel execution time in seconds.
We repeated each experiment five times to ensure robustness and minimize measurement noise. 
In two sets of experiments, we test two versions of \gcsnapC: 
A. \gcsnapC based on mpi4py.futures and B. \gcsnapC based on mpi4py.futures with improved taxonomy information parsing.
In a third set of experiments C. we also evaluate the portability and performance of B. with many input targets. Finally, we present the execution profile of C.

It is worth noting that GCsnap's workflow Step 3 (Annotate) requires annotation files, which are not readily available.
Therefore, in all our experiments with \gcsnapC, we disabled the majority of annotations to ensure a fair comparison with \gcsnapD.
Consequently, the sole substeps executed in Step 3 are: 
(1) annotation of families with PDB structures, (2) clustering of operons, and (3) generation of the phylogenetic tree based on taxonomy data.

\begin{table*}[!hb]
    \small  
    \centering
    \caption{Design of factorial experiments to evaluate the performance of \gcsnapC.}
    \vspace{-2.5mm}
    \label{tab:experiments}
    \begin{tblr}{width=\textwidth, colspec={Q[l,m,wd=20mm]|Q[l,m,wd=55mm]|Q[l,m,wd=77mm]},stretch=1.2} 
        \textbf{Factor} & \textbf{Value}\textsuperscript{a}  & \textbf{Properties} \\ 
        \hline      
        \SetCell[r=3]{m} Experiment Set
        & \SetCell[r=1]{m} \makecell[l]{\textcolor{myred}{A.: \gcsnapC~with mpi4py.futures}\textsuperscript{b}} &
        \makecell[l]{Number of targets: 1'000;\\ Number of computing nodes: 4, 8, 16; \\
        Number of MPI ranks per node: 2, 4, 8, 16; \\ Number of CPU cores per rank: 1, 2, 4;} \\ 
        \cline[dashed]{2-3}            
        & \SetCell[r=1]{m} \makecell[l]{\textcolor{myred}{B.: A. + Improved taxonomy parsing}\textsuperscript{c}} &
        \makecell[l]{Number of targets: 1'000;\\ Number of computing nodes: 4, 8, 16; \\
        Number of MPI ranks per node: 2, 4, 8, 16; \\ Number of CPU cores per rank: 1, 2, 4;} \\
        \cline[dashed]{2-3}    
        & \SetCell[r=1]{m} \makecell[l]{\textcolor{myblue}{C.: Portability and Performance of B.}\textsuperscript{c}} &
        \makecell[l]{Number of targets: 10'000;\\ Number of computing nodes: 1; \\
        Number of MPI ranks per node: 4, 8; \\ Number of CPU cores per rank: 8, 16, 32;} \\      
        \cline[dashed]{2-3}    
        & \SetCell[r=1]{m} \makecell[l]{\textcolor{myblue}{Profiling of C.}\textsuperscript{d}} &
        \makecell[l]{Number of targets: 10'000;\\ Number of computing nodes: 1; \\
        Number of MPI ranks per node: 8; \\ Number of CPU cores per rank: 8;} \\   
        \hline          
        \SetCell[r=2]{m} \makecell[l]{Computing\\system} &
        \SetCell[r=1]{m} \textcolor{myred}{Xeon nodes} &  \makecell[l]{\textcolor{myred}{2 CPU Xeon E5-2640v4, 2.4GHz, each with 10 cores} \\ \textcolor{myred}{and 25 MB L3 cache;}\\ \textcolor{myred}{64 GB RAM; CentOS Linux release 7.9.2009; }} \\ 
        \cline[dashed]{2-3}  
        & \SetCell[r=1]{m} \textcolor{myblue}{AMD node} &  \makecell[l]{\textcolor{myblue}{2 AMD EPYC 7742, 2.25 GHz,  each with 64 cores} \\ \textcolor{myblue}{and 256 MB L3 cache;} \\ \textcolor{myblue}{1'500 GB RAM; CentOS Linux release 7.9.2009;}} \\          
        \hline
        Experiment & Repetitions & 5 \\ 
        \hline
        Metric & \makecell[l]{Parallel execution time} & 
        \makecell[l]{Average time per experiment across all repetitions (seconds)} \\  
        \hline 
    \end{tblr}
    \begin{tablenotes}
        \item \hspace{-12pt} \textsuperscript{a} Color indicates on which computing system the experiment set was executed.
        \item \hspace{-12pt} \textsuperscript{b} \url{https://github.com/GCsnap/gcsnap2cluster/tree/unperforming-taxonomy}
        \item \hspace{-12pt} \textsuperscript{c} \url{https://github.com/GCsnap/gcsnap2cluster/tree/main}
        \item \hspace{-12pt} \textsuperscript{d} \url{https://github.com/GCsnap/gcsnap2cluster/tree/profiling}
    \end{tablenotes}
\end{table*}  

\begin{figure}[t]
    \centering
    \begin{subfigure}{\linewidth} 
        \centering
        \includegraphics[width=0.6\linewidth]{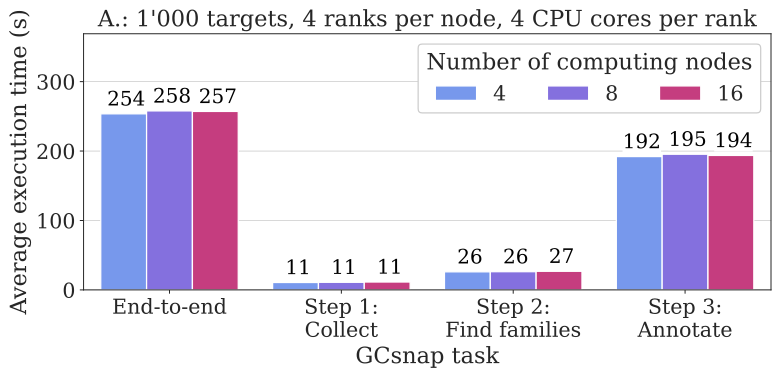} 
        \caption{\gcsnapC{} with mpi4py.futures and different numbers of computing nodes, 1'000 input targets, 4 MPI ranks per node, and 4 CPU cores per rank.}
        \label{fig:gcsnap2c_steps_1}
    \end{subfigure}
    \begin{subfigure}{\linewidth} 
        \centering
        \includegraphics[width=0.6\linewidth]{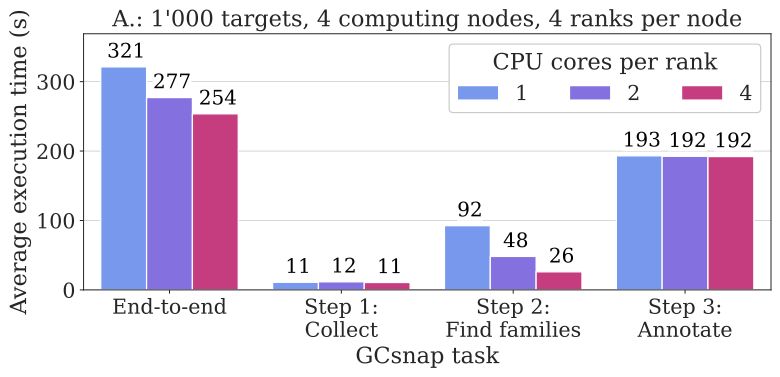} 
        \caption{\gcsnapC{} with mpi4py.futures and different numbers of CPU cores per rank, 1'000 input targets, 4 computing nodes, and 4 MPI ranks per node.}
        \label{fig:gcsnap2c_steps_2}
    \end{subfigure}
    \begin{subfigure}{\linewidth}
        \centering
        \includegraphics[width=0.6\linewidth]{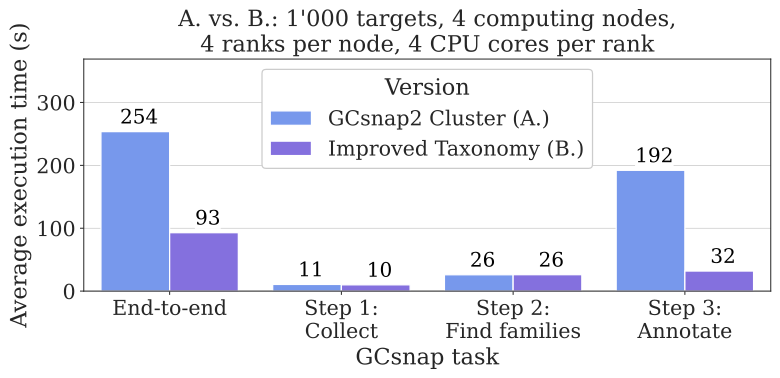} 
        \caption{\gcsnapC{} with mpi4py.futures and improved taxonomy parsing with 1'000 input targets, 4 computing nodes, 4 MPI ranks per node, and 4 CPU cores per rank.}
        \label{fig:gcsnap2c_steps_3}
    \end{subfigure}
    \caption{Average end-to-end and step-wise parallel execution time of \gcsnapC across experiment sets A and B and over 5 repetitions, with various numbers of computing nodes, MPI ranks per node, and CPU cores per rank. One MPI rank holds one Python process created with mpi4py.futures, each with as many OpenMP threads for MMseqs2 as there are CPU cores per rank.
    Experiments conducted on various nodes with 2 Xeon E5-2640v4 CPUs and 20 CPU cores.}
    \label{fig:gcsnap2c_steps}
\end{figure}

\subsection{A. \gcsnapC with mpi4py.futures} 
Figure~\ref{fig:gcsnap2c_steps} shows the results of executing the entire workflow of \gcsnapC from Figure~\ref{fig:modularization_cluster} with mpi4py.futures on 1'000 target sequences with 4 MPI ranks per node and a varying number of computing nodes and CPU cores per rank.
A notable decrease in the overall execution time is evident when comparing \gcsnapC to \gcsnapD. 
Analyzing the genomic context with the latter took 166 seconds for 100 targets. In contrast, the new implementation processed 10× more target sequences on average in 256 seconds, as shown in Figure~\ref{fig:gcsnap2c_steps_1}, decreasing the per target execution time from 1.66 to 0.256 seconds.
Conversely, the end-to-end execution time remains unaffected by increasing the number of computing nodes. 
There is no improvement in Step 1 and a slight performance degradation for Step 3 with an increasing number of nodes.

In contrast, changing the number of CPU cores has a very significant influence on the average execution time of Step 2, as illustrated in Figure~\ref{fig:gcsnap2c_steps_2}.
Quadrupling the number of CPU cores per rank reduces the average execution time from 92 seconds to 26 seconds. The reason is that MMseqs2, which is used to compute all-against-all sequence similarities, supports multi-threading through OpenMP~\cite{MMseqs_paper} and hence, benefits from additional CPU cores per rank. 
This is the only part of \gcsnapC that uses explicit multithreading. 
Certain Python modules, such as Pandas~\cite{pandas}, contain functionality that implicitly benefits from additional cores by running multi-threaded without GIL-imposed restriction by design.
Apart from those two cases, parallelism in \gcsnapC is exploited through processes via the mpi4py.futures module, where each Python process corresponds to one MPI rank. 

Of concern regarding the results shown in Figures~\ref{fig:gcsnap2c_steps_1} and \ref{fig:gcsnap2c_steps_2} is the significant execution time spent on Step 3. This indicates potential for optimization, which we describe next.

\subsection{B. Improved Taxonomy Parsing}
\label{subs:improved_taxonomy_parsing}
The \gcsnapC implementation using mpi4py.futures uses Python dictionaries for taxonomy parsing.
Python dictionaries running on multicore environments are bound to the GIL, which limits parallelism and performance of taxonomy parsing.
To circumvent this GIL-imposed constraint and improve processing speeds, we incorporate Pandas data frames.
Figure~\ref{fig:gcsnap2c_steps_3} illustrates the execution time for the Pandas-based variant.
Clearly, the improved taxonomy handling results in a notable enhancement in end-to-end performance.
Overall, the analysis of 1'000 targets with 4 nodes and 4 ranks per node is executed in 93 seconds, averaging 0.093 seconds per target.

\subsection{C. Portability and Performance of \gcsnapC with mpi4py.futures and Improved Taxonomy Parsing}
\begin{figure}[t]
    \centering 
    \includegraphics[width=0.6\linewidth]{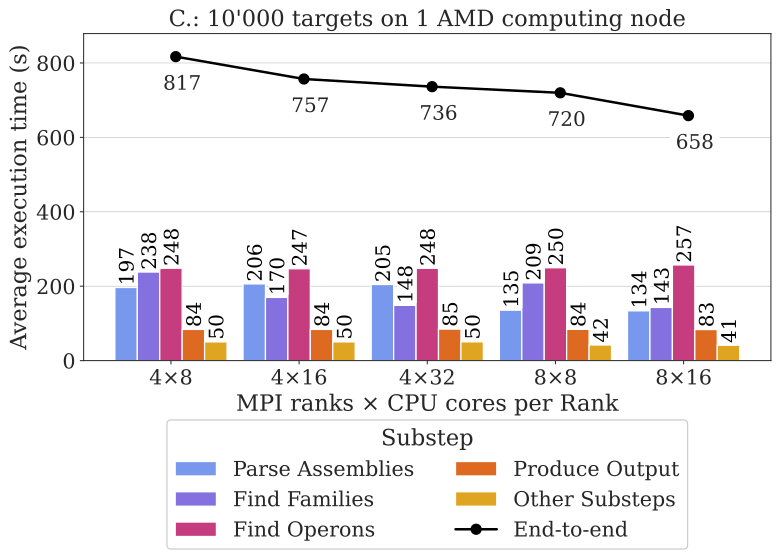}
    \caption{Average execution time of individual substeps of \gcsnapC over 5 repetitions with different numbers of MPI ranks × CPU cores and 10'000 input targets. One MPI rank holds one Python process created with mpi4py.futures, each with as many OpenMP threads for MMseqs2 as there are CPU cores per rank.
    Experiments were conducted on 1 computing node with two AMD EPYC 7742 CPUs, each with 64 CPU cores.}
    \label{fig:gcsnap2c_scale}
\end{figure}

To assess our claim that \gcsnapC with mpi4py.futures and the improved taxonomy parsing can handle large workloads effectively, we executed it with 10'000 target sequences.
Based on previous results of limited scalability with an increasing number of nodes (Figure~\ref{fig:gcsnap2c_steps_1}), but increased performance with increasing number of CPU cores (Figure~\ref{fig:gcsnap2c_steps_2}), we decided to conduct this experiment on an AMD compute node with 128 CPU cores. 
The results are presented in Figure~\ref{fig:gcsnap2c_scale}.
The figure shows individual substeps rather than complete workflow steps: parsing assemblies (a substep of Step 1), identifying families (from Step 2), finding operons (from Step 3), and producing the output (beyond Step 3), which for 10'000 targets accounts for a significant amount of the overall execution time. 
Other individual substeps not explicitly displayed are grouped and reported under "Other Substeps".

Several findings are worth emphasizing:
First, the average end-to-end execution time is around 740 seconds for 10'000 sequences, equating to a processing time of 0.074 seconds per target.
In contrast to 1.66 seconds per target with \gcsnapD,  \gcsnapC is 22× faster. 

Second, \gcsnapC can analyze 10'000 sequences in a reasonable time, allowing for previously infeasible large-scale genomic context analyses.

Third, the initial results (Figure~\ref{fig:gcsnap2c_steps_2}) with the \textit{find families} substep performing better with more CPU cores per MPI rank are confirmed. 
The reason is the computation of all-against-all sequence similarities with MMseqs2, which profits from an increased number of CPU cores due to exploiting multicore parallelism via OpenMP multithreading. 
This substep also benefits from additional MPI processes. 
Keeping the number of CPU cores per MPI rank fixed at 8 while increasing the number of MPI ranks from 4 to 8 results in a reduction in the execution time of this substep from 238 to 209 seconds. 
A detailed explanation of the cause follows in Section~\ref{sec:profiling}.

Fourth, \gcsnapC driven by mpi4py.futures also works well in a single-node environment, not only in multi-node environments, proving its flexibility in terms of resources used for execution.

However, parts of \gcsnapC do not scale well. 
Increasing the number of MPI ranks from 4 to 8 with 8 CPU cores per MPI rank only reduces the time needed to parse 10'000 assemblies from 197 to 135 seconds.
Moreover, the long execution time for finding operons indicates that the current implementation of \gcsnapC could benefit from further optimization, planned for future work.

\subsection{Profiling Analysis of \gcsnapD and \gcsnapC} \label{sec:profiling}
\begin{figure}[t]
    \centering 
    \includegraphics[width=0.6\linewidth]{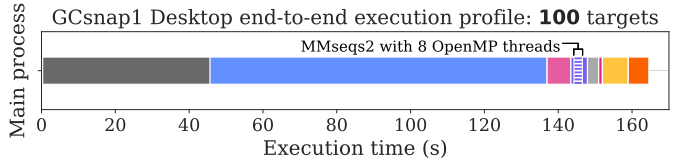}   
    \includegraphics[width=0.6\linewidth]{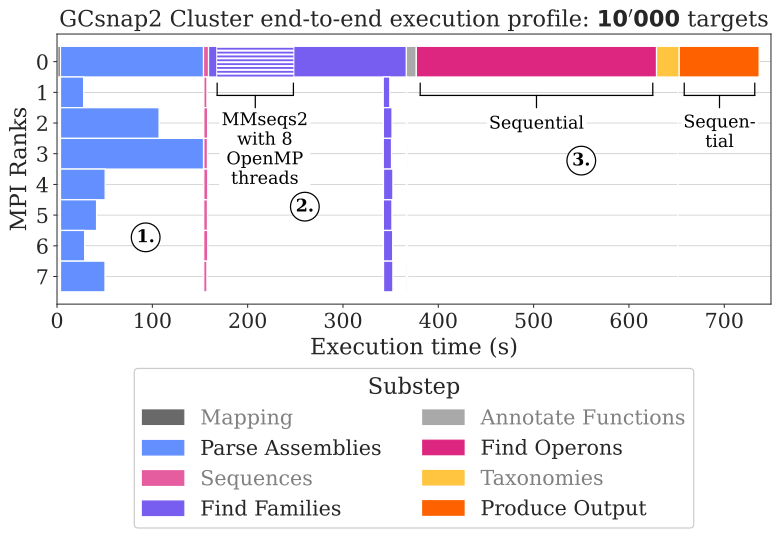}
    \caption{Execution profiles of \gcsnapD and \gcsnapC. 
    Top: \gcsnapD with 100 targets and executing with one process on a node with 8 CPU cores, requiring 166 seconds (as shown in Figure~\ref{fig:gcsnap1_overview}).    
    Bottom: \gcsnapC with 10'000 targets, and executing on one node with 8 MPI ranks and 8 CPU cores per rank. 
    One MPI rank holds one Python process created with mpi4py.futures, each with as many OpenMP threads for MMseqs2 as there are CPU cores per rank. 
    Experiments were conducted on 1 computing node with 2$\times$ AMD EPYC 7742 CPUs, each with 64 CPU cores. 
    The substeps \textcolor{mygrey}{mapping}, \textcolor{lightmygrey}{annotate functions}, and \textcolor{myyellow}{taxonomies} are in fact MPI parallel, but the execution time of ranks 1 to 7 is very short, such that they are not visible on the end-to-end execution scale of 720 seconds (reported in Figure~\ref{fig:gcsnap2c_scale}).    }
    \label{fig:profile}
\end{figure}

The poor performance of \gcsnapD can be attributed to the Collect workflow Step 1 as described in Section~\ref{sec:assessment}.
More insight into which substep limits the performance of \gcsnapD can be found through profiling.
Using the measured execution times for each substep with 100 targets and 8 OpenMP threads on 8 CPU cores (see Figure~\ref{fig:gcsnap1_overview}), we created the execution profile shown in Figure~\ref{fig:profile}.

Clearly, the mapping of target sequence IDs and the parsing of assemblies are the most time-consuming substeps of the end-to-end execution time. 
The single process version of \gcsnapD cannot take advantage of the multithread-based execution used in Python due to the restriction imposed by the GIL, with the exception of the \textit{find families} substep. 
As mentioned above, this substep uses MMseqs2 which employs OpenMP for multithreaded execution, shown as lines within the main Python process.

Understanding the limitations of \gcsnapC described in the previous section involves profiling the code and its MPI execution pattern. 
Performance analysis tools such as \emph{score-P} with its Python bindings \cite{scoreP, scorePpython} or \emph{mpiP} \cite{mpiP} are designed to collect information about MPI ranks. 
However, both require the presence of explicit MPI API calls in the code (here, Python), and most importantly \emph{MPI.Init}.
With mpi4py.futures, no explicit MPI API calls are required, rendering the profiling of \gcsnapC with these tools impossible.
Therefore, to create a profile of the execution performance, we adapted the parallel wrapper function\footnote{\url{https://github.com/GCsnap/gcsnap2cluster/tree/profiling}} to collect the necessary data from each MPI rank.
Together with the timestamps for each execution step contained in the detailed log files generated during a \gcsnapC run, these data can be used to create a parallel execution profile with 8 MPI ranks and 8 CPU cores per rank (see Figure~\ref{fig:gcsnap2c_scale}) as shown in Figure~\ref{fig:profile} for \gcsnapC.

The main finding from the profiling-based performance analysis is that the proportion of substeps using MPI-based parallelism is lower than expected. 
\gcsnapC is inherently sequential, where a substep spends most time in integrating the results from the MPI ranks into the data collected or computed in previous substeps.
The part that benefits most from MPI-based parallelism is the parsing of assembly files \circled{1.}, although not as efficiently as possible. 
It seems that this substep suffers from load imbalance caused by our static workload distribution scheme. 
By design, there are as many chunks of workload as there are MPI workers involved, which is not suitable when the workload is heterogeneous among chunks. 
As listed in Table~\ref{tab:data}, the size of the assembly file varies widely (from a few KB to hundreds of MB), resulting in an imbalanced distribution between ranks.
This substep would benefit from a more fine-grained and/or dynamic workload distribution of chunks. The latter is possible with Python as shown in \cite{mendeh}.

Another interesting finding concerns the \textit{find families} substep \circled{2.}. 
As mentioned above, MMseqs2 uses OpenMP threads indicated by lines within the coordinator MPI rank. 
The part that is actually MPI parallel is the assignment of the identified families to the target sequences and their neighboring genes.
The limiting factor in this substep is the compute-bound hierarchical clustering from SciPy, which is not parallel.

A limitation concerns the \textit{find operons} and \textit{produce output} substeps of Step \circled{3.}. 
An early design decision was to implement these without relying on mpi4py.futures, as they were not considered a bottleneck based on the profile of \gcsnapD (Figure~\ref{fig:profile}--top). 
Clearly, for large workloads with 10'000 target sequences, our initial assessment was misleading.
Based on this observation, our efforts will include the parallelization of these two substeps to further reduce the end-to-end execution time of \gcsnapC.

\subsection{Discussion} \label{sec:discussion}
The results of this study highlight the significant progress made with \gcsnapC in overcoming the limitations of \gcsnapD. 
The comparative assessment of Dask.jobqueue and mpi4py.futures showed that mpi4py.futures is a better choice for distributed execution due to its superior performance when running on more than two computing nodes. 
Through redesign and leveraging mpi4py.futures for distributed execution, \gcsnapC achieved a 22× improvement in end-to-end execution time and the capability to handle much larger datasets. 
These advances enable genomic studies that were previously infeasible due to computational and implementation constraints. 

Despite this progress, several challenges still remain.
An important limitation is the reliance on pre-downloaded data, as presented in Section~\ref{sec:data}, which requires frequent updates to maintain accuracy and relevance. 
Moreover, with the resolved I/O bottleneck of \gcsnapD, the computation has now become the primary limiting factor in \gcsnapC.
Certain aspects of our implementation remain unoptimized, particularly the workload distribution for \textit{parse assembly} in workflow Step 1 and the substep for \textit{find operons} of workflow Step 3.
This suggests that further improvement of these modules is necessary through algorithmic refinements to obtain additional performance gains. 

Our experiments explored two execution configurations for \gcsnapC:
many nodes with a few CPU cores each (Figure~\ref{fig:gcsnap2c_steps}) and
a single node with a high number of CPU cores (Figure~\ref{fig:gcsnap2c_scale}).
What remains to be analyzed is the scalability of \gcsnapC across multiple nodes with high(er) core counts. 
A logical next step is to combine these experiments to evaluate {\gcsnapC}’s performance in such distributed, high-core environments.

An issue that remains insufficiently addressed is the functional annotation of families and their members (the first substep in workflow Step 3).
Currently, this substep depends on user input, as the required data is not readily available for download.
A future goal is to develop a more automated approach to eliminate the need for users to prepare such files manually in advance.

Notwithstanding these limitations, \gcsnapC shows a substantial advancement over its predecessor \gcsnapD.
The methodology for developing \gcsnapC not only addresses scalability issues in genomic context studies, but also demonstrates its wider potential to solve similar challenges in other scientific applications that increasingly rely on large-scale data analysis.

\section{Related Work} \label{sec:rel_work}
Over the last decade, several tools with similar functionality to GCsnap have been developed. For example, Locus Visualization (LoVis4u) is a Python package that can be integrated into existing analysis pipelines, also providing a command-line interface~\cite{Egorov2024}. 
The resulting output is vector graphics of publication quality. 
Similarly to GCsnap, LoVis4u depends on MMseqs2~\cite{MMseqs_paper} to determine sequence similarities and cluster flanking genes into putative conserved protein families. 
The input sequences must originate from GenBank. 
This differs from GCsnap, which can process a multitude of input identifiers from a variety of protein ID standards.

Another tool is FlaGs~\cite{Saha2020}, which enables users to create visualizations for genomic context analysis. 
The supported input identifiers are limited to those from RefSeq, which differs from GCsnap because it uses assemblies from both the RefSeq and GenBank databases. 
FlaGs employs the Jackhammer method, which is based on Hidden Markov Models, to identify clusters among the flanking genes. 
In contrast to GCsnap, the generated visualizations lack interactivity. 
However, FlaGs offers both a web-based version and the option to run it locally. 
In the study by Ernits et al.~\cite{Ernits2023} a variant of FlaGs, designated NetFlax, was used to examine proteinaceous toxin-antitoxin (TA) systems.

Another representative of web-based tools is GeCoViz~\cite{botas2022}, which facilitates the creation of interactive visualizations of the genomic context of prokaryotes alone. 
GCsnap, on the other hand, is capable of handling sequences derived from both prokaryotic and eukaryotic organisms. 
GeCoViz employs a precomputed information storage system, namely a MongoDB database, to facilitate rapid search results. 

Another Web-based tool is the Microbial Genomic Context Viewer (MGcV)~\cite{Overmars2013}, which is, however, limited to prokaryotes. 
Similarly, a MySQL database is used in the background, with updates occurring on a weekly basis. 
The distinction lies in the nature of the stored data. 
In contrast to GeCoViz, which stores precalculated information, MGcV stores the data necessary for the calculation, ensuring rapid access. 

In addition, there is the web-based Genomic Context Viewer (GCV) version 2, which employs a distinct methodology~\cite{Cleary2023}. 
This tool is intended to facilitate the visualization of microsyntenies, which are defined as small genomic regions derived from a common ancestor. 
GCV is capable of analyzing relationships in real-time through on-demand computation across federated data sources. 
Furthermore, it is possible to integrate GCV into existing web applications.

Several studies underscore the versatility and efficiency of parallel Python and mpi4py in addressing computationally-intensive problems in various domains. 
Ye et al.~\cite{Ye2023} used mpi4py to compute meshes from the East and Gulf Coast for composite flooding simulations, where parallelization enabled continental-scale processing. 
Similarly, mpi4py and NumPy were used to simulate hybrid particle-field molecular dynamics~\cite{Ledum2023}. 
In addition, Szustak et al.~\cite{Szustak2023} demonstrated the benefits of mpi4py in social sciences with task-parallel social network simulations and data-parallel kernel polynomial methods (KPM) for graph eigenvalue computations. 
An example using mpi4py.futures is ParaEMT~\cite{xiong2024}, an electromagnetic transient simulation tool designed to execute on HPC systems.

\textit{To our knowledge, \gcsnapC is the first scalable application for genomic context analysis in the field of bioinformatics parallelized with mpi4py}.

\section{Conclusion} \label{sec:conclusion}
This work introduces \gcsnapC, a scalable and user-friendly tool for genomic context analysis, designed to overcome the limitations of its predecessor \gcsnapD, which could not handle large inputs. 
{\gcsnapC}'s design features a modular architecture and the flexibility to execute in various computational environments, enabling bioinformatics analyses of hundreds of thousands of input sequences in a matter of several hours, while also supporting the development of custom workflows.

\gcsnapC can exploit hardware parallelism by combining multiple MPI processes through mpi4py.futures and multiple threads for MMseqs2 execution with OpenMP.
However, as this work has shown, specific approaches are needed to exploit the inherent parallelism available in each substep under programming and dataflow constraints.

Future developments of \gcsnapC will focus on streamlining data update processes, maintaining accessibility, and its ease of use for life scientists.
This includes substantial data downloads and database preparations in advance.
Regular updates are crucial to ensure the accuracy and relevance of the information. 
Therefore, a promising solution is to implement a fully automated data pipeline to streamline and standardize these processes.

\section*{Acknowledgments}
We thank the members of the Ciorba and Schwede groups at the University of Basel for their valuable support, as well as the support team of the National Institute of Health for their assistance with NCBI data sources. This work was partially supported by
the University of Basel Research Fund for Excellent Junior Researchers (Grant number: U.570.0006 to J.P.), 
Swiss National Science Foundation Weave project (Grant number: 310030L\textunderscore220141 to T.S.), 
SIB Swiss Institute of Bioinformatics, 
PASC (Swiss Platform for Advanced Scientific Computing) project ”SPH-EXA2: Optimizing Smoothed Particle Hydrodynamics for Exascale Computing” (to O.S.S and F.M.C.) and
SERI-funded SKACH consortium (to O.S.S and F.M.C.). 

\section*{Open Research and Reproducibility}
The implementation of \gcsnapC is publicly available on GitHub: \url{https://github.com/GCsnap/gcsnap2cluster}, v1.0.0.
As the project is actively under development, future versions will include additional features and improvements.
In addition to the source code and build instructions, the \gcsnapC repository contains scripts to download the raw data and create the required databases.
The version used to assess the execution time of \gcsnapD can be accessed via the branch named "timing" at: \url{https://github.com/RetoKrummenacher/GCsnap}.
All material used for this paper is available on Zenodo~\cite{Zenodo}, including files containing the targets, scripts to run the experiments, results, plots and Conda environment files.

\bibliographystyle{ACM-Reference-Format}
\bibliography{references}


\begin{thebibliography}{33}


\ifx \showCODEN    \undefined \def \showCODEN     #1{\unskip}     \fi
\ifx \showDOI      \undefined \def \showDOI       #1{#1}\fi
\ifx \showISBNx    \undefined \def \showISBNx     #1{\unskip}     \fi
\ifx \showISBNxiii \undefined \def \showISBNxiii  #1{\unskip}     \fi
\ifx \showISSN     \undefined \def \showISSN      #1{\unskip}     \fi
\ifx \showLCCN     \undefined \def \showLCCN      #1{\unskip}     \fi
\ifx \shownote     \undefined \def \shownote      #1{#1}          \fi
\ifx \showarticletitle \undefined \def \showarticletitle #1{#1}   \fi
\ifx \showURL      \undefined \def \showURL       {\relax}        \fi
\providecommand\bibfield[2]{#2}
\providecommand\bibinfo[2]{#2}
\providecommand\natexlab[1]{#1}
\providecommand\showeprint[2][]{arXiv:#2}

\bibitem[Botas et~al\mbox{.}(2022)]%
        {botas2022}
\bibfield{author}{\bibinfo{person}{Jorge Botas}, \bibinfo{person}{{\'A}lvaro Rodríguez~del Río}, \bibinfo{person}{Joaquín Giner-Lamia}, {and} \bibinfo{person}{Jaime Huerta-Cepas}.} \bibinfo{year}{2022}\natexlab{}.
\newblock \showarticletitle{GeCoViz: genomic context visualisation of prokaryotic genes from a functional and evolutionary perspective}.
\newblock \bibinfo{journal}{\emph{Nucleic Acids Research}} \bibinfo{volume}{50}, \bibinfo{number}{W1} (\bibinfo{date}{May} \bibinfo{year}{2022}), \bibinfo{pages}{W352--W357}.
\newblock
\showISSN{1362-4962}
\urldef\tempurl%
\url{https://doi.org/10.1093/nar/gkac367}
\showDOI{\tempurl}


\bibitem[Chaudhari et~al\mbox{.}(2024)]%
        {chaudhari_biological_2024}
\bibfield{author}{\bibinfo{person}{Jyoti~Kant Chaudhari}, \bibinfo{person}{Shubham Pant}, \bibinfo{person}{Richa Jha}, \bibinfo{person}{Rajesh~Kumar Pathak}, {and} \bibinfo{person}{Dev~Bukhsh Singh}.} \bibinfo{year}{2024}\natexlab{}.
\newblock \showarticletitle{Biological big-data sources, problems of storage, computational issues, and applications: a comprehensive review}.
\newblock \bibinfo{journal}{\emph{Knowledge and Information Systems}} \bibinfo{volume}{66}, \bibinfo{number}{6} (\bibinfo{date}{June} \bibinfo{year}{2024}), \bibinfo{pages}{3159--3209}.
\newblock
\showISSN{0219-3116}
\urldef\tempurl%
\url{https://doi.org/10.1007/s10115-023-02049-4}
\showDOI{\tempurl}


\bibitem[Cleary and Farmer(2023)]%
        {Cleary2023}
\bibfield{author}{\bibinfo{person}{Alan~M Cleary} {and} \bibinfo{person}{Andrew~D Farmer}.} \bibinfo{year}{2023}\natexlab{}.
\newblock \showarticletitle{Genome Context Viewer (GCV) version 2: enhanced visual exploration of multiple annotated genomes}.
\newblock \bibinfo{journal}{\emph{Nucleic Acids Research}} \bibinfo{volume}{51}, \bibinfo{number}{W1} (\bibinfo{date}{May} \bibinfo{year}{2023}), \bibinfo{pages}{W225--W231}.
\newblock
\showISSN{1362-4962}
\urldef\tempurl%
\url{https://doi.org/10.1093/nar/gkad391}
\showDOI{\tempurl}


\bibitem[community(2024)]%
        {scipy}
\bibfield{author}{\bibinfo{person}{The~SciPy community}.} \bibinfo{year}{2024}\natexlab{}.
\newblock \bibinfo{title}{scipy.cluster.hierarchy.linkage}.
\newblock
\newblock
\urldef\tempurl%
\url{https://docs.scipy.org/doc/scipy/reference/generated/scipy.cluster.hierarchy.linkage.html}
\showURL{%
\tempurl}


\bibitem[Dalcin(2024)]%
        {mpi4pyfutures}
\bibfield{author}{\bibinfo{person}{Lisandro Dalcin}.} \bibinfo{year}{2024}\natexlab{}.
\newblock \bibinfo{title}{mpi4py.futures}.
\newblock
\newblock
\urldef\tempurl%
\url{https://mpi4py.readthedocs.io/en/stable/mpi4py.futures.html}
\showURL{%
Retrieved November 28, 2024 from \tempurl}


\bibitem[Dalcin and Fang(2021)]%
        {Dalcin2021}
\bibfield{author}{\bibinfo{person}{Lisandro Dalcin} {and} \bibinfo{person}{Yao-Lung~L. Fang}.} \bibinfo{year}{2021}\natexlab{}.
\newblock \showarticletitle{mpi4py: Status Update After 12 Years of Development}.
\newblock \bibinfo{journal}{\emph{Computing in Science \& Engineering}} \bibinfo{volume}{23}, \bibinfo{number}{4} (\bibinfo{year}{2021}), \bibinfo{pages}{47--54}.
\newblock
\urldef\tempurl%
\url{https://doi.org/10.1109/MCSE.2021.3083216}
\showDOI{\tempurl}


\bibitem[Egorov and Atkinson(2024)]%
        {Egorov2024}
\bibfield{author}{\bibinfo{person}{Artyom~A. Egorov} {and} \bibinfo{person}{Gemma~C. Atkinson}.} \bibinfo{year}{2024}\natexlab{}.
\newblock \showarticletitle{LoVis4u: Locus Visualisation tool for comparative genomics}.
\newblock \bibinfo{journal}{\emph{bioRxiv}} (\bibinfo{year}{2024}), \bibinfo{numpages}{6}~pages.
\newblock
\urldef\tempurl%
\url{https://doi.org/10.1101/2024.09.11.612399}
\showDOI{\tempurl}


\bibitem[Ernits et~al\mbox{.}(2023)]%
        {Ernits2023}
\bibfield{author}{\bibinfo{person}{Karin Ernits}, \bibinfo{person}{Chayan~Kumar Saha}, \bibinfo{person}{Tetiana Brodiazhenko}, \bibinfo{person}{Bhanu Chouhan}, \bibinfo{person}{Aditi Shenoy}, \bibinfo{person}{Jessica~A. Buttress}, \bibinfo{person}{Julián~J. Duque-Pedraza}, \bibinfo{person}{Veda Bojar}, \bibinfo{person}{Jose~A. Nakamoto}, \bibinfo{person}{Tatsuaki Kurata}, \bibinfo{person}{Artyom~A. Egorov}, \bibinfo{person}{Lena Shyrokova}, \bibinfo{person}{Marcus J.~O. Johansson}, \bibinfo{person}{Toomas Mets}, \bibinfo{person}{Aytan Rustamova}, \bibinfo{person}{Jelisaveta Džigurski}, \bibinfo{person}{Tanel Tenson}, \bibinfo{person}{Abel Garcia-Pino}, \bibinfo{person}{Henrik Strahl}, \bibinfo{person}{Arne Elofsson}, \bibinfo{person}{Vasili Hauryliuk}, {and} \bibinfo{person}{Gemma~C. Atkinson}.} \bibinfo{year}{2023}\natexlab{}.
\newblock \showarticletitle{The structural basis of hyperpromiscuity in a core combinatorial network of type II toxin–antitoxin and related phage defense systems}.
\newblock \bibinfo{journal}{\emph{Proceedings of the National Academy of Sciences}} \bibinfo{volume}{120}, \bibinfo{number}{33}, Article \bibinfo{articleno}{e2305393120} (\bibinfo{date}{Aug.} \bibinfo{year}{2023}), \bibinfo{numpages}{12}~pages.
\newblock
\showISSN{1091-6490}
\urldef\tempurl%
\url{https://doi.org/10.1073/pnas.2305393120}
\showDOI{\tempurl}


\bibitem[for Biotechnology~Information(2024)]%
        {Blastp}
\bibfield{author}{\bibinfo{person}{NCBI: National~Center for Biotechnology~Information}.} \bibinfo{year}{2024}\natexlab{}.
\newblock \bibinfo{title}{Basic Local Alignment Search Tool}.
\newblock
\newblock
\urldef\tempurl%
\url{https://blast.ncbi.nlm.nih.gov/Blast.cgi}
\showURL{%
Retrieved December 6, 2024 from \tempurl}


\bibitem[Forum(2023)]%
        {mpi}
\bibfield{author}{\bibinfo{person}{Message Passing~Interface Forum}.} \bibinfo{year}{2023}\natexlab{}.
\newblock \bibinfo{title}{MPI: A Message-Passing Interface Standard}.
\newblock
\newblock
\urldef\tempurl%
\url{https://www.mpi-forum.org/docs/mpi-4.1/mpi41-report.pdf}
\showURL{%
Retrieved November 27, 2024 from \tempurl}


\bibitem[Foundation(2024)]%
        {sqlite}
\bibfield{author}{\bibinfo{person}{Python~Software Foundation}.} \bibinfo{year}{2024}\natexlab{}.
\newblock \bibinfo{title}{sqlite3 — DB-API 2.0 interface for SQLite databases}.
\newblock
\newblock
\urldef\tempurl%
\url{https://docs.python.org/3/library/sqlite3.html}
\showURL{%
\tempurl}


\bibitem[Inc. and Contributors(2018)]%
        {dask}
\bibfield{author}{\bibinfo{person}{Anaconda Inc.} {and} \bibinfo{person}{Contributors}.} \bibinfo{year}{2018}\natexlab{}.
\newblock \bibinfo{title}{Dask: A Python library for parallel and distributed computing}.
\newblock
\newblock
\urldef\tempurl%
\url{https://docs.dask.org/en/stable/}
\showURL{%
Retrieved November 28, 2024 from \tempurl}


\bibitem[Index(2024)]%
        {pandas}
\bibfield{author}{\bibinfo{person}{Python~Package Index}.} \bibinfo{year}{2024}\natexlab{}.
\newblock \bibinfo{title}{Pandas}.
\newblock
\newblock
\urldef\tempurl%
\url{https://pypi.org/project/pandas/}
\showURL{%
Retrieved December 6, 2024 from \tempurl}


\bibitem[Krummenacher et~al\mbox{.}(2025)]%
        {Zenodo}
\bibfield{author}{\bibinfo{person}{Reto Krummenacher}, \bibinfo{person}{Osman~S. Simsek}, \bibinfo{person}{Michèle Leemann}, \bibinfo{person}{Leila~T. Alexander}, \bibinfo{person}{Torsten Schwede}, \bibinfo{person}{Florina~M. Ciorba}, {and} \bibinfo{person}{Joana Pereira}.} \bibinfo{year}{2025}\natexlab{}.
\newblock \showarticletitle{Artifacts for Scalable Genomic Context Analysis with GCsnap2 on HPC Clusters}.
\newblock \bibinfo{journal}{\emph{Zenodo}} (\bibinfo{year}{2025}).
\newblock
\urldef\tempurl%
\url{https://doi.org/10.5281/zenodo.15301856}
\showDOI{\tempurl}


\bibitem[Laboratory(2020)]%
        {mpiP}
\bibfield{author}{\bibinfo{person}{Lawrence Livermore~National Laboratory}.} \bibinfo{year}{2020}\natexlab{}.
\newblock \bibinfo{title}{mpiP}.
\newblock
\newblock
\urldef\tempurl%
\url{https://github.com/LLNL/mpiP}
\showURL{%
Retrieved March 10, 2025 from \tempurl}


\bibitem[Ledum et~al\mbox{.}(2023)]%
        {Ledum2023}
\bibfield{author}{\bibinfo{person}{Morten Ledum}, \bibinfo{person}{Manuel Carrer}, \bibinfo{person}{Samiran Sen}, \bibinfo{person}{Xinmeng Li}, \bibinfo{person}{Michele Cascella}, {and} \bibinfo{person}{Sigbjørn~Løland Bore}.} \bibinfo{year}{2023}\natexlab{}.
\newblock \showarticletitle{HylleraasMD: Massively parallel hybrid particle-field molecular dynamics in Python}.
\newblock \bibinfo{journal}{\emph{Journal of Open Source Software}} \bibinfo{volume}{8}, \bibinfo{number}{84}, Article \bibinfo{articleno}{4149} (\bibinfo{year}{2023}), \bibinfo{numpages}{4}~pages.
\newblock
\urldef\tempurl%
\url{https://doi.org/10.21105/joss.04149}
\showDOI{\tempurl}


\bibitem[Mavromatis et~al\mbox{.}(2009)]%
        {Mavromatis2009}
\bibfield{author}{\bibinfo{person}{Konstantinos Mavromatis}, \bibinfo{person}{Ken Chu}, \bibinfo{person}{Natalia Ivanova}, \bibinfo{person}{Sean~D. Hooper}, \bibinfo{person}{Victor~M. Markowitz}, {and} \bibinfo{person}{Nikos~C. Kyrpides}.} \bibinfo{year}{2009}\natexlab{}.
\newblock \showarticletitle{Gene Context Analysis in the Integrated Microbial Genomes (IMG) Data Management System}.
\newblock \bibinfo{journal}{\emph{PLoS ONE}} \bibinfo{volume}{4}, \bibinfo{number}{11}, Article \bibinfo{articleno}{e7979} (\bibinfo{date}{Nov.} \bibinfo{year}{2009}), \bibinfo{numpages}{8}~pages.
\newblock
\showISSN{1932-6203}
\urldef\tempurl%
\url{https://doi.org/10.1371/journal.pone.0007979}
\showDOI{\tempurl}


\bibitem[Mendhe(2021)]%
        {mendeh}
\bibfield{author}{\bibinfo{person}{Abhilash Mendhe}.} \bibinfo{year}{2021}\natexlab{}.
\newblock \bibinfo{title}{Dynamic loop self-scheduling on distributed-memory systems with Python}.
\newblock
\newblock
\urldef\tempurl%
\url{https://hpc.dmi.unibas.ch/scientific_output/completed_theses_and_projects/}
\showURL{%
Retrieved March 16, 2025 from \tempurl}
\newblock
\shownote{CSP.14}.


\bibitem[Müllner(2011)]%
        {muellner}
\bibfield{author}{\bibinfo{person}{Daniel Müllner}.} \bibinfo{year}{2011}\natexlab{}.
\newblock \bibinfo{title}{Modern hierarchical, agglomerative clustering algorithms}.
\newblock
\newblock
\showeprint[arxiv]{1109.2378}
\urldef\tempurl%
\url{https://arxiv.org/abs/1109.2378}
\showURL{%
\tempurl}


\bibitem[of~Bioinformatics(2023)]%
        {atlas}
\bibfield{author}{\bibinfo{person}{SIB Swiss~Institute of Bioinformatics}.} \bibinfo{year}{2023}\natexlab{}.
\newblock \bibinfo{title}{Protein Universe Atlas}.
\newblock
\newblock
\urldef\tempurl%
\url{https://uniprot3d.org/}
\showURL{%
Retrieved December 2, 2024 from \tempurl}


\bibitem[Overmars et~al\mbox{.}(2013)]%
        {Overmars2013}
\bibfield{author}{\bibinfo{person}{Lex Overmars}, \bibinfo{person}{Robert Kerkhoven}, \bibinfo{person}{Roland~J Siezen}, {and} \bibinfo{person}{Christof Francke}.} \bibinfo{year}{2013}\natexlab{}.
\newblock \showarticletitle{MGcV: the microbial genomic context viewer for comparative genome analysis}.
\newblock \bibinfo{journal}{\emph{BMC Genomics}}  \bibinfo{volume}{14}, Article \bibinfo{articleno}{209} (\bibinfo{year}{2013}), \bibinfo{numpages}{9}~pages.
\newblock
\showISSN{1471-2164}
\urldef\tempurl%
\url{https://doi.org/10.1186/1471-2164-14-209}
\showDOI{\tempurl}


\bibitem[Pereira(2021)]%
        {Pereira2021}
\bibfield{author}{\bibinfo{person}{Joana Pereira}.} \bibinfo{year}{2021}\natexlab{}.
\newblock \showarticletitle{GCsnap: Interactive Snapshots for the Comparison of Protein-Coding Genomic Contexts}.
\newblock \bibinfo{journal}{\emph{Journal of Molecular Biology}} \bibinfo{volume}{433}, \bibinfo{number}{11}, Article \bibinfo{articleno}{166943} (\bibinfo{date}{May} \bibinfo{year}{2021}), \bibinfo{numpages}{8}~pages.
\newblock
\showISSN{0022-2836}
\urldef\tempurl%
\url{https://doi.org/10.1016/j.jmb.2021.166943}
\showDOI{\tempurl}


\bibitem[Rocklin(2015)]%
        {Rocklin2015}
\bibfield{author}{\bibinfo{person}{Matthew Rocklin}.} \bibinfo{year}{2015}\natexlab{}.
\newblock \showarticletitle{Dask: Parallel computation with Blocked algorithms and Task Scheduling}. In \bibinfo{booktitle}{\emph{Proceedings of the 14th Python in Science Conference}}. \bibinfo{pages}{126--132}.
\newblock
\urldef\tempurl%
\url{https://doi.org/10.25080/Majora-7b98e3ed-013}
\showDOI{\tempurl}


\bibitem[Rogowski et~al\mbox{.}(2023)]%
        {rogowski2023}
\bibfield{author}{\bibinfo{person}{Marcin Rogowski}, \bibinfo{person}{Samar Aseeri}, \bibinfo{person}{David Keyes}, {and} \bibinfo{person}{Lisandro Dalcin}.} \bibinfo{year}{2023}\natexlab{}.
\newblock \showarticletitle{mpi4py.futures: MPI-Based Asynchronous Task Execution for Python}.
\newblock \bibinfo{journal}{\emph{IEEE Transactions on Parallel \& Distributed Systems}} \bibinfo{volume}{34}, \bibinfo{number}{02} (\bibinfo{date}{Feb.} \bibinfo{year}{2023}), \bibinfo{pages}{611--622}.
\newblock
\urldef\tempurl%
\url{https://doi.org/10.1109/TPDS.2022.3225481}
\showDOI{\tempurl}


\bibitem[Saha et~al\mbox{.}(2020)]%
        {Saha2020}
\bibfield{author}{\bibinfo{person}{Chayan~Kumar Saha}, \bibinfo{person}{Rodrigo Sanches~Pires}, \bibinfo{person}{Harald Brolin}, \bibinfo{person}{Maxence Delannoy}, {and} \bibinfo{person}{Gemma~Catherine Atkinson}.} \bibinfo{year}{2020}\natexlab{}.
\newblock \showarticletitle{FlaGs and webFlaGs: discovering novel biology through the analysis of gene neighbourhood conservation}.
\newblock \bibinfo{journal}{\emph{Bioinformatics}} \bibinfo{volume}{37}, \bibinfo{number}{9} (\bibinfo{date}{Dec.} \bibinfo{year}{2020}), \bibinfo{pages}{1312--1314}.
\newblock
\showISSN{1367-4811}
\urldef\tempurl%
\url{https://doi.org/10.1093/bioinformatics/btaa788}
\showDOI{\tempurl}


\bibitem[(SBC)({[n.\,d.]})]%
        {phobius}
\bibfield{author}{\bibinfo{person}{Stockholm Bioinformatics~Center (SBC)}.} \bibinfo{year}{[n.\,d.]}\natexlab{}.
\newblock \bibinfo{title}{Phobius: A combined transmembrane topology and signal peptide predictor}.
\newblock
\newblock
\urldef\tempurl%
\url{https://phobius.sbc.su.se/}
\showURL{%
Retrieved December 6, 2024 from \tempurl}


\bibitem[Score-P(2023)]%
        {scorePpython}
\bibfield{author}{\bibinfo{person}{Score-P}.} \bibinfo{year}{2023}\natexlab{}.
\newblock \bibinfo{title}{Score-P Python Bindings}.
\newblock
\newblock
\urldef\tempurl%
\url{https://github.com/score-p/scorep_binding_python}
\showURL{%
Retrieved March 10, 2025 from \tempurl}


\bibitem[Steinegger and Söding(2018)]%
        {MMseqs_paper}
\bibfield{author}{\bibinfo{person}{Martin Steinegger} {and} \bibinfo{person}{Johannes Söding}.} \bibinfo{year}{2018}\natexlab{}.
\newblock \showarticletitle{Clustering huge protein sequence sets in linear time}.
\newblock \bibinfo{journal}{\emph{Nature Communications}}  \bibinfo{volume}{9}, Article \bibinfo{articleno}{2542} (\bibinfo{date}{June} \bibinfo{year}{2018}), \bibinfo{numpages}{8}~pages.
\newblock
\urldef\tempurl%
\url{https://doi.org/10.1038/s41467-018-04964-5}
\showDOI{\tempurl}


\bibitem[Szustak et~al\mbox{.}(2023)]%
        {Szustak2023}
\bibfield{author}{\bibinfo{person}{Lukasz Szustak}, \bibinfo{person}{Marcin Lawenda}, \bibinfo{person}{Sebastian Arming}, \bibinfo{person}{Gregor Bankhamer}, \bibinfo{person}{Christoph Schweimer}, {and} \bibinfo{person}{Robert Elsässer}.} \bibinfo{year}{2023}\natexlab{}.
\newblock \showarticletitle{Profiling and optimization of Python-based social sciences applications on HPC systems by means of task and data parallelism}.
\newblock \bibinfo{journal}{\emph{Future Generation Computer Systems}}  \bibinfo{volume}{148} (\bibinfo{year}{2023}), \bibinfo{pages}{623--635}.
\newblock
\showISSN{0167-739X}
\urldef\tempurl%
\url{https://doi.org/10.1016/j.future.2023.07.005}
\showDOI{\tempurl}


\bibitem[Tech({[n.\,d.]})]%
        {tmhmm}
\bibfield{author}{\bibinfo{person}{Danmarks Tekniske Universitet (DTU)~Health Tech}.} \bibinfo{year}{[n.\,d.]}\natexlab{}.
\newblock \bibinfo{title}{TMHMM-2.0}.
\newblock
\newblock
\urldef\tempurl%
\url{https://services.healthtech.dtu.dk/services/TMHMM-2.0/}
\showURL{%
Retrieved December 6, 2024 from \tempurl}


\bibitem[{Virtual Institute — High Productivity Supercomputing}(2025)]%
        {scoreP}
\bibfield{author}{\bibinfo{person}{{Virtual Institute — High Productivity Supercomputing}}.} \bibinfo{year}{2025}\natexlab{}.
\newblock \bibinfo{title}{Score-P}.
\newblock
\newblock
\urldef\tempurl%
\url{https://www.vi-hps.org/projects/score-p}
\showURL{%
Retrieved March 10, 2025 from \tempurl}


\bibitem[Xiong et~al\mbox{.}(2024)]%
        {xiong2024}
\bibfield{author}{\bibinfo{person}{Min Xiong}, \bibinfo{person}{Bin Wang}, \bibinfo{person}{Deepthi Vaidhynathan}, \bibinfo{person}{Jonathan Maack}, \bibinfo{person}{Matthew~J. Reynolds}, \bibinfo{person}{Andy Hoke}, \bibinfo{person}{Kai Sun}, {and} \bibinfo{person}{Jin Tan}.} \bibinfo{year}{2024}\natexlab{}.
\newblock \showarticletitle{ParaEMT: An Open Source, Parallelizable, and HPC-Compatible EMT Simulator for Large-Scale IBR-Rich Power Grids}.
\newblock \bibinfo{journal}{\emph{IEEE Transactions on Power Delivery}} \bibinfo{volume}{39}, \bibinfo{number}{2} (\bibinfo{year}{2024}), \bibinfo{pages}{911--921}.
\newblock
\urldef\tempurl%
\url{https://doi.org/10.1109/TPWRD.2023.3342715}
\showDOI{\tempurl}


\bibitem[Ye et~al\mbox{.}(2023)]%
        {Ye2023}
\bibfield{author}{\bibinfo{person}{Fei Ye}, \bibinfo{person}{Linlin Cui}, \bibinfo{person}{Yinglong Zhang}, \bibinfo{person}{Zhengui Wang}, \bibinfo{person}{Saeed Moghimi}, \bibinfo{person}{Edward Myers}, \bibinfo{person}{Greg Seroka}, \bibinfo{person}{Alan Zundel}, \bibinfo{person}{Soroosh Mani}, {and} \bibinfo{person}{John~G.W. Kelley}.} \bibinfo{year}{2023}\natexlab{}.
\newblock \showarticletitle{A parallel Python-based tool for meshing watershed rivers at continental scale}.
\newblock \bibinfo{journal}{\emph{Environmental Modelling \& Software}}  \bibinfo{volume}{166}, Article \bibinfo{articleno}{105731} (\bibinfo{year}{2023}), \bibinfo{numpages}{14}~pages.
\newblock
\showISSN{1364-8152}
\urldef\tempurl%
\url{https://doi.org/10.1016/j.envsoft.2023.105731}
\showDOI{\tempurl}


\end{thebibliography}
\end{document}